\documentclass[10pt,a4paper,twocolumn,USenglish,prb,aps,floatfix,groupedaddress,superscriptaddress,longbibliography]{revtex4-2}

\usepackage{amsmath}
\usepackage{amssymb}
\usepackage{amsfonts}
\usepackage{dcolumn} 
\usepackage{bm}
\usepackage{bbm}
\usepackage{color,array}
\usepackage{colortbl}
\usepackage{dsfont}

\usepackage[USenglish]{babel}
\usepackage[T1]{fontenc}
\usepackage[utf8]{inputenc}
\usepackage{graphicx}
\usepackage{upgreek}
\usepackage{mathtools}
\usepackage[unicode]{hyperref}
\usepackage[dvipsnames]{xcolor}

\hypersetup{
	colorlinks   = true, 
	urlcolor     = blue, 
	linkcolor    = blue, 
	citecolor   = blue 
}

\def\wR{\omega_{\mathrm{R}}}
\def\muB{\mu_\mathrm{B}}
\def\TR{T_{\rm R}}
\def\Bext{B_\mathrm{ext}}
\def\ez{\bm{e}_z}
\def\tausg{\tau_\mathrm{s}}
\def\tauseff{\tau_\mathrm{s}^*}
\def\wn{\omega_\mathrm{N}}
\def\TR{T_\mathrm{R}}
\def\fm{f_\mathrm{m}}

\begin{document}
	
\title{Resonant spin amplification in Faraday geometry}
		
\date{\today}
	
\author{P.~Schering}
\affiliation{Condensed Matter Theory, Technische Universit\"at Dortmund, 44221 Dortmund, Germany}

\author{E.~Evers}
\affiliation{Experimentelle Physik 2, Technische Universit\"at Dortmund, 44221 Dortmund, Germany}

\author{V.~Nedelea}
\affiliation{Experimentelle Physik 2, Technische Universit\"at Dortmund, 44221 Dortmund, Germany}
	
\author{D.~S.~Smirnov}
\affiliation{Ioffe Institute, Russian Academy of Sciences, 194021 St.\,Petersburg, Russia}

\author{E.~A.~Zhukov}
\affiliation{Experimentelle Physik 2, Technische Universit\"at Dortmund, 44221 Dortmund, Germany}
\affiliation{Ioffe Institute, Russian Academy of Sciences, 194021 St.\,Petersburg, Russia}

\author{D.~R.~Yakovlev}
\affiliation{Experimentelle Physik 2, Technische Universit\"at Dortmund, 44221 Dortmund, Germany}
\affiliation{Ioffe Institute, Russian Academy of Sciences, 194021 St.\,Petersburg, Russia}

\author{M.~Bayer}
\affiliation{Experimentelle Physik 2, Technische Universit\"at Dortmund, 44221 Dortmund, Germany}
\affiliation{Ioffe Institute, Russian Academy of Sciences, 194021 St.\,Petersburg, Russia}

\author{G.~S.~Uhrig}
\affiliation{Condensed Matter Theory, Technische Universit\"at Dortmund, 44221 Dortmund, Germany}

\author{A.~Greilich}
\affiliation{Experimentelle Physik 2, Technische Universit\"at Dortmund, 44221 Dortmund, Germany}

\begin{abstract}
We demonstrate the realization of the resonant spin amplification~(RSA) effect in Faraday geometry where a magnetic field is applied parallel to the optically induced spin polarization so that no RSA is expected. However, model considerations predict that it can be realized for a central spin interacting with a fluctuating spin environment. As a demonstrator, we choose an ensemble of singly-charged (In,Ga)As/GaAs quantum dots, where the resident electron spins interact with the surrounding nuclear spins. The observation of RSA in Faraday geometry requires intense pump pulses with a high repetition rate and can be enhanced by means of the spin-inertia effect. Potentially, it provides the most direct and reliable tool to measure the longitudinal $g$~factor of the charge carriers.
\end{abstract}

\maketitle

The possibility of using the spin degree of freedom for quantum information~\cite{NielsenChuang,Ladd2010} continues to drive research on semiconductor nanostructures~\cite{Loss1998,Gangloff2019,Denning2019,Chekhovich2020}. The main characteristic in this field is defined by the lifetime of the information or the spin coherence time. 
Complementarily, the development of spintronics~\cite{Hirohata2020} over two decades gave birth to a plethora of experimental tools for the investigation of the spin dynamics in semiconductor nanostructures. A major part of these methods is based on the interrelation between the spin of a charge carrier and the polarization of a photon emitted or absorbed by the semiconductor structure~\cite{OptOr}. The most popular ones are the Hanle effect~\cite{Hanle1924} and the time-resolved pump-probe technique, based on the pulsed-laser excitation~\cite{AwschalomBook02,dyakonov_book,slavcheva_book}, which can be extended to detect the spin dynamics on arbitrary long timescales with femtosecond resolution~\cite{Extended_pp}. Other powerful tools are the spin-noise spectroscopy~\cite{Sinitsyn2016,Smirnov2020_review} and the spin-inertia technique~\cite{Heisterkamp2015,Zhukov2018,Smirnov2018,Schering2019}. 

One of the most basic parameters of the spin dynamics is the $g$~factor, which is often anisotropic in semiconductor nanostructures. Its transverse component can be measured very precisely when a magnetic field is applied in Voigt geometry by means of the resonant spin amplification~(RSA) effect~\cite{Kikkawa1998}. It is based on the pump-probe technique, where the spin polarization is measured at a fixed pump-probe delay as a function of a transverse magnetic field. Provided the spin relaxation time is longer than the laser repetition period, the spin polarization is amplified when the Larmor precession period is a multiple integer of the laser repetition period~\cite{AwschalomBook02}. The RSA effect can also be used to evaluate the spin relaxation time, the spread of the transverse $g$~factor, and the strength of the hyperfine interaction~\cite{Glazov2008}. The RSA method has been successfully applied to a variety of systems, ranging from bulk GaAs~\cite{Kikkawa1998}, III-V and II-VI~quantum wells and epilayers~\cite{YugovaPRL09,Zhukov2012,greilich2012ZnSe:F}, to quantum~dots~\cite{Varwig2012,Yugova2012}.

In Faraday geometry where the magnetic field is parallel to the optical axis, there is no spin precession of the charge carriers on average such that it is difficult to determine their $g$~factor.
Up to now, it must be measured indirectly: First, the transverse $g$~factor is determined in Voigt geometry, e.g., by standard RSA~\cite{Kikkawa1998} or from time-resolved measurement of quantum beats~\cite{Jeune1997,Greilich2006a,Yugova2007}. Repeating the measurements in an oblique geometry (e.g., tilted by $45^\circ$) gives access to the longitudinal $g$~factor~\cite{Jeune1997,Zhukov2012}.
In this Letter, we propose a RSA-based method to measure the longitudinal $g$~factor \emph{directly} and with high accuracy. More precisely, we demonstrate that RSA can emerge in Faraday geometry for an ensemble of $n$-doped (In,Ga)As/GaAs quantum dots~(QDs). The effect is enabled by the hyperfine interaction between the resident electron spins and the fluctuating nuclear spin environment~\cite{Schering2019}.

\paragraph*{Experimental details --}
\label{sec:experimental-details}

The studied sample consists of 20 layers of (In,Ga)As QDs separated by $70$-nm barriers of GaAs and grown by molecular beam epitaxy on a \mbox{(100)-oriented} GaAs substrate. A $\updelta$-doping layer of silicon $16\,$nm above each QD layer provides a single electron per QD on average. Rapid thermal annealing at $880\,^{\circ}$C for $30\,$s homogenizes the QD size distribution and shifts the average emission energy to $1.3662\,$eV. The QD density per layer amounts to $10^{10}\,\text{cm}^{-2}$.

\begin{figure}[t]
	\begin{center}
		\includegraphics[trim=0mm 0mm 0mm 0mm, clip, width=\columnwidth]{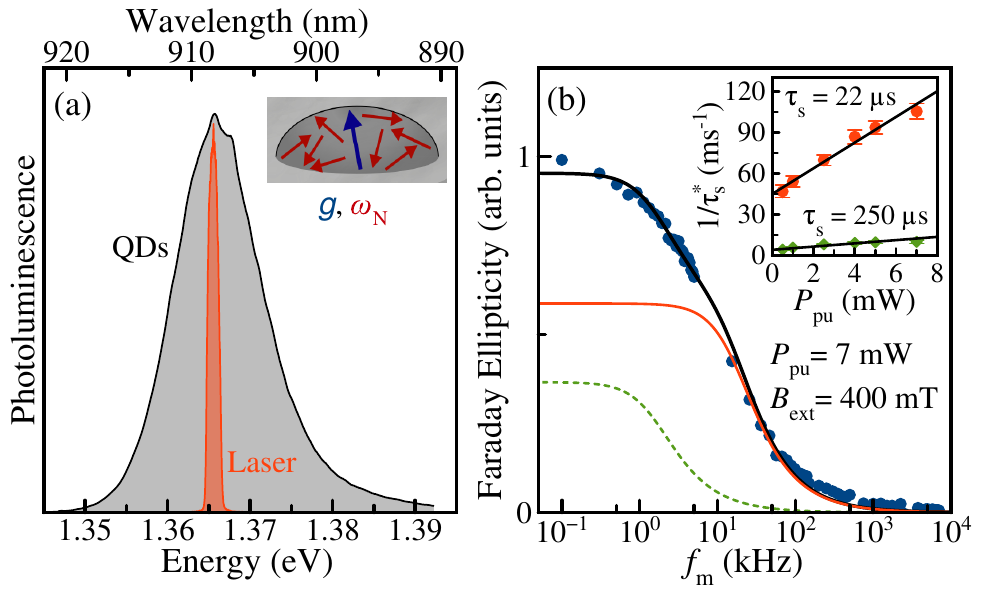}
		\caption{(a)~Photoluminescence of the QD ensemble (gray) at \mbox{$T=5.3\,$K} along with the spectrum of the laser used in the pump-probe measurements (orange). The sketch shows an \mbox{$n$-doped} QD containing an electron spin (blue) with a $g$~factor, which is interacting with the fluctuating nuclear spin environment characterized by the frequency~$\wn$ (red).
		(b)~Faraday ellipticity amplitude as a function of the pump modulation frequency $\fm$ for a pump power of $P_\mathrm{pu} = 7\,$mW measured at a magnetic field of $\Bext = 400\,$mT with a pump-probe delay of $-60\,$ps (blue circles). The black line is the two-component fit, see main text. The contributions of each component are shown by the solid-orange and dashed-green lines. The inset shows the power dependence of the two corresponding inverse effective spin lifetimes $1/\tauseff$. A linear extrapolation to zero power (black lines) yields $\tausg=(22 \pm 1)\,\mu$s (orange) and $\tausg=(250 \pm 27)\,\mu$s~(green).}
		\label{fig1}
	\end{center}
\end{figure}

The sample is cooled to $5.3\,$K in a Helium gas atmosphere inside a cryostat with a split-coil magnet. A superconducting solenoid pair creates an external magnetic field~$\Bext \ez$ in the direction of light incidence, i.e., along the optical $z$~axis with an accuracy of 2 degree (Faraday geometry). The sample is illuminated by laser pulses with a central optical energy of $1.3655\,$eV and a full width at half maximum of $1.3\,$meV. The pulses have a duration of $2\,$ps and are emitted with a repetition frequency of $1\,$GHz. They are split in pump and probe pulses, which are degenerate in photon energy and shifted by $0.7$\,meV to the low-energy flank of the QD photoluminescence, see Fig.~\ref{fig1}(a). The pump pulses are directed along a variable mechanical delay line. A double-modulation scheme reduces the noise arising from separate detection of the scattered pump and probe light. The helicity of the pump is modulated by an electro-optical modulator at a frequency $\fm$ between left- and right-handed circular polarization ranging from $0.1$~to~$10^4\,$kHz, preventing the build-up of significant dynamic nuclear polarization~\cite{OptOr_Chap5}. The linearly polarized probe beam is intensity modulated using a photoelastic modulator at a frequency of $100\,$kHz in series with a Glan prism. For signal detection, the reference frequency of the lock-in amplifier is running at the difference frequency. The pump and the probe beams are focused on the same sample spot, with the pump focused to a spot diameter of $50\,\mu$m and the probe to a diameter of $45\,\mu$m. We measure the Faraday ellipticity amplitude of the probe pulses using an optical polarization bridge, which consists of a lambda-quarter plate, a Wollaston prism, and a balanced photodetector. The Faraday ellipticity is proportional to the electron spin polarization of the QDs along the optical axis~\cite{Yugova2009}.

\paragraph*{Results --}
\label{sec:results}

Each circularly polarized pump pulse partially orients the spins of the resident electrons along the optical axis~\cite{Shabaev2003}. The spin polarization added by each pulse in this way depends on its effective pulse area~$\Theta$~\cite{Greilich2006a}, which is determined by the average pump beam power and scales like $P_\mathrm{pu} \propto \Theta^2$ when the power is small~\cite{Yugova2009}. If the added spin polarization exceeds the relaxed polarization along the optical axis until the next pulse, the spin polarization builds up. This is the case for a magnetic field of \mbox{$\Bext=400\,$mT} using a pump power of \mbox{$P_\mathrm{pu} = 7\,$mW}, and the origin for the spin polarization in Fig.~\ref{fig1}(b) displayed by the Faraday ellipticity at $-60$\,ps time delay between pump and probe pulses studied as a function of the pump modulation frequency~$\fm$. The decay of the ellipticity upon increasing $\fm$ stems from the spin-inertia effect~\cite{Heisterkamp2015,Zhukov2018,Smirnov2018,Schering2019}. It can be described by the dependence \mbox{$E(\fm) = E_0/\sqrt{1+(2\pi \fm \tauseff)^2}$}, where $\tauseff$ is the effective spin lifetime at the corresponding pump power. The extrapolation of $\tauseff$ to zero power allows for extracting the intrinsic spin relaxation $\tausg$ of the electrons~\cite{Heisterkamp2015,Schering2019}. The inset in Fig.~\ref{fig1}(b) depicts the extracted times for two components present in the spin-inertia dependence shown as solid-red and dashed-green curves. We relate them to two effective subsets of electrons in the QD ensemble and focus on the regime $\fm \geq 5\,$kHz for which only the shorter living subensemble with $\tausg = (22 \pm 1)\,\mu$s contributes significantly~\footnote{Probably, there is a continuous distribution of spin relaxation times due to the inhomogeneous character of the QD ensemble. However, this is difficult to describe by our model where only a single spin relaxation time enters for the ground state~\cite{Supplement}. In any case, a single-component fit of the spin-inertia dependence is much worse than the two-component fit which captures all effects important for the following analysis.}.

\begin{figure}[t]
	\begin{center}
		\includegraphics[trim=0mm 0mm 0mm 0mm, clip, width=\columnwidth]{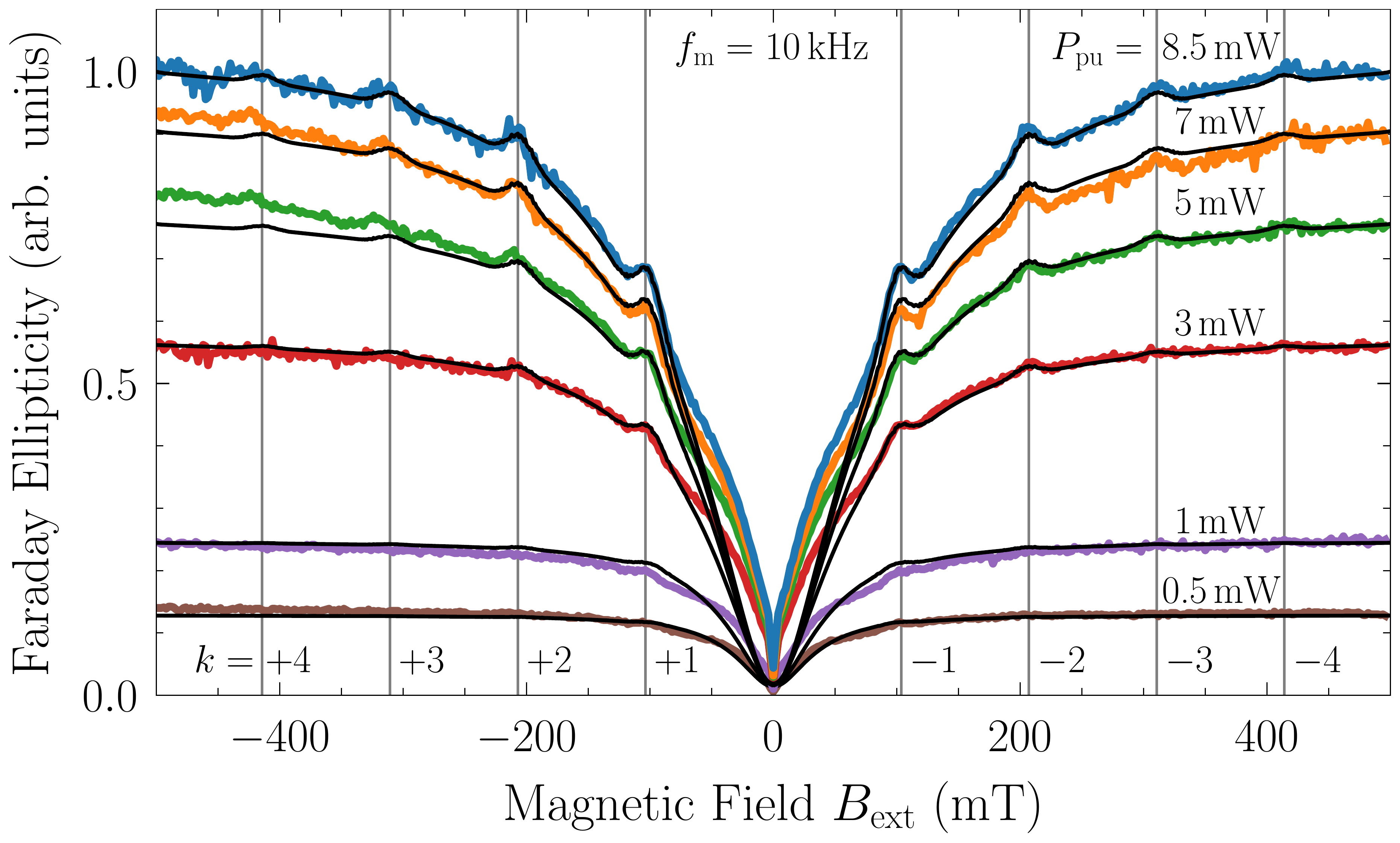}
		\caption{PRCs for different pump powers given next to the curves at a pump modulation frequency of \mbox{$\fm = 10\,$kHz}. The colored data is the experimentally measured Faraday ellipticity, the smooth black curves are the simulated PRCs. The pulse area in the simulations~\cite{Supplement} is chosen to fit the experimentally obtained power dependence of the Faraday ellipticity at large field. The positions of the small peaks visible in the PRCs for larger pump powers match the $k$th~RSA mode (vertical grey lines) as predicted by the PSC~\eqref{PSC}, yielding $g_z = -0.69 \pm 0.01$ for the longitudinal electronic $g$~factor.}
		\label{fig2}
	\end{center}
\end{figure}

\begin{figure}[t]
	\begin{center}
		\includegraphics[trim=0mm 0mm 0mm 0mm, clip, width=\columnwidth]{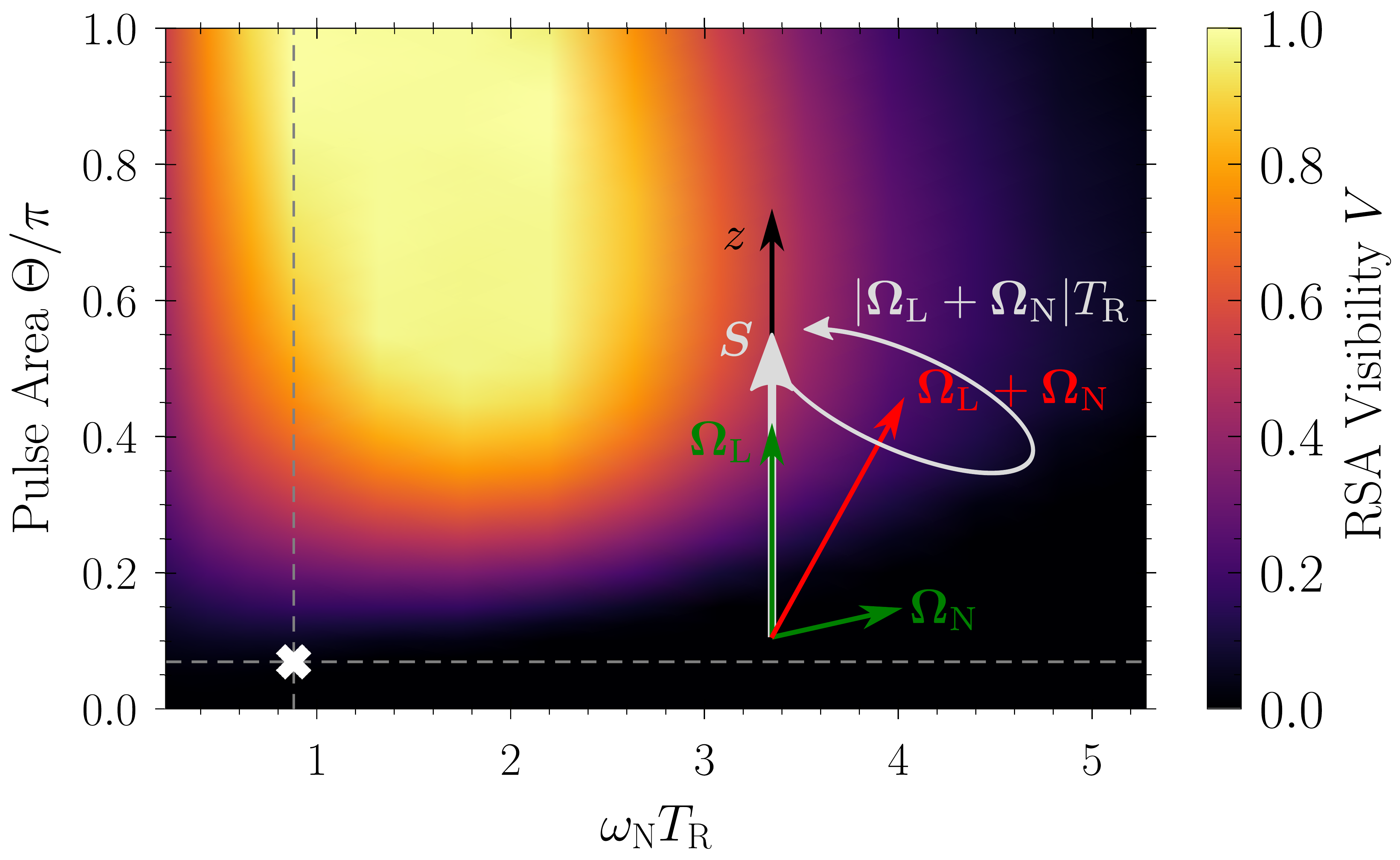}
		\caption{Visibility map for the RSA mode $|k| = 1$ in Faraday geometry modeled for detection by Faraday ellipticity. The pump-probe delay is set to zero, the pump modulation frequency to $\fm = 10\,$kHz. The white cross marks the experimental conditions for a pump power of $P_\mathrm{pu} = 8.5\,$mW ($\Theta = 0.07\pi$) and a pulse repetition period of \mbox{$\TR = 1\,$ns} [$\wn/(2\pi) = 140\,$MHz]. The sketch illustrates the mechanism leading to RSA in an external longitudinal magnetic field~$\bm\Omega_\mathrm{L}$~\cite{Schering2019}. The transverse components of the Overhauser field~$\bm\Omega_\mathrm{N}$ tilt the resulting effective magnetic field from the $z$~axis, inducing a precession of the electron spin~$\bm S$ about the effective field $\bm\Omega_\mathrm{eff} = \bm\Omega_\mathrm{L} + \bm\Omega_\mathrm{N}$ with frequency~$\Omega_\mathrm{eff}$.}
		\label{fig3}
	\end{center}
\end{figure}

Without a magnetic field, the spin polarization of the electrons decays due to the hyperfine interaction with the nuclear spin fluctuations~\cite{Merkulov2002}. The application of a longitudinal magnetic field suppresses the nuclei-induced spin relaxation and in turn increases the spin polarization~\cite{Smirnov2020b}. This effect is referred to as polarization recovery.
For the $n$-doped QD sample studied in this work, the polarization recovery curves~(PRCs) have the typical $\mathsf{V}$-like shape (symmetric around zero field)~\cite{Petrov2008,Zhukov2018,Smirnov2018,Schering2019,Smirnov2020b}. 
Figure~\ref{fig2} shows the PRCs for a wide range of pump powers. For small powers, the electron spin polarization is minimal at $0\,$mT and rises to a saturation level within $100\,$mT. For zero field, the electron spin is only subject to the isotropic nuclear fluctuation field characterized by the frequency~\mbox{$\wn/(2\pi) = 140\,$MHz}~\footnote{The value $\wn/(2\pi) = 140\,$MHz is estimated by fitting our model~\cite{Supplement} to the measured PRCs. The width of the zero-field minimum does not only depend on $\wn$ but also on other parameters due to saturation effects~\cite{Schering2019}. Furthermore, there are also contributions from resident or photoexcited holes as we will discuss later. Hence, we mainly focused on fitting to the width of the later introduced RSA modes, which appears to be robust against variations of the other parameters.}
(equivalent to a field of~$14.5\,$mT), which is the characteristic frequency of the electronic Larmor precession in the nuclear fluctuation field (Overhauser field) leading to nuclei-induced spin relaxation; see the sketch in Fig.~\ref{fig1}(a). 
Upon increase of the longitudinal magnetic field, the role of the Overhauser field is reduced due to the increase of the electronic Zeeman splitting. In a classical picture, the amplitude of the electron spin precession about the resulting effective magnetic field (superposition of Overhauser and external field; see the sketch in Fig.~\ref{fig3}) is reduced the more the larger the external field. Hence, the lifetime of the spin polarization and therefore the polarization itself increases until saturation is reached~\cite{Glazov_book}.

For larger pump powers, the spin polarization increases and the qualitative dependence on the magnetic field is very similar. But strong pulses suppress the in-plane electron spin components, effectively accelerating the spin relaxation, which in turn leads to a broadening of the zero-field minimum in the PRC~\cite{Schering2019,Smirnov2020b}.
Clearly, this is the case in Fig.~\ref{fig2}: The larger the pump power, the larger the required magnetic field to reach saturation.

The appearance of a modulation in the Faraday ellipticity at certain values of the longitudinal magnetic field in the case of large pump powers is most striking.
This is the result of RSA in Faraday geometry enabled by nuclear fluctuation fields~\cite{Schering2019}. As demonstrated in Fig.~\ref{fig2}, the modulations appear at magnetic fields~$\Bext$ which fulfill the phase synchronization condition~(PSC)
\begin{align}
	\Omega_\mathrm{L} = |k| \wR \,, \qquad k \in \mathds{Z} \,, \label{PSC}
\end{align}
for the Larmor frequency $\Omega_\mathrm{L} = \muB |g_z \Bext| \hbar^{-1}$ ($\muB$~is the Bohr magneton and $\hbar$~the reduced Planck constant).
These discrete resonance frequencies are given by multiples of the laser repetition frequency~$\wR$. We label them by the mode number~$k$.
The longitudinal electronic $g$~factor~$|g_z|$ determines the mode positions.
While we cannot extract its sign from this effect, we know that it is negative for similar (In,Ga)As/GaAs QDs~\cite{Yugova2007}.
Taking all peak positions into account, we obtain $g_z = -0.69 \pm 0.01$. For comparison, the transverse $g$~factor of the electrons in this sample amounts to $g_\perp = -0.599 \pm 0.001$~\footnote{The transverse $g$~factor $g_\perp = -0.599 \pm 0.001$ is determined from time-resolved measurements of the spin dynamics in a transverse magnetic field of $2\,$T.}.

The theoretical model is presented in the Supplemental Material~\cite{Supplement}. It almost perfectly describes the experiment as shown by the black lines in Fig.~\ref{fig2}. Noteworthy, in contrast to Refs.~\cite{Yugova2009,Smirnov2018,Schering2019}, the model accounts for a lifetime of the photoexcited trion which is comparable to the pulse repetition period of $1\,$ns.
The main deviation between experiment and theory is found in the regime of very small magnetic fields. This deviation has a narrow $\mathsf{M}$-like shape, which is typical for $p$-doped QD samples with a strong field dependent spin generation rate~\cite{Zhukov2018,Smirnov2018}. Hence, we attribute the deviation to resident or photoexcited hole spins with weak hyperfine interaction.

In what follows, we describe how the nuclear fluctuation fields in QDs enable us to implement RSA in Faraday geometry.
In a single QD, the localized electron spin~$\bm S$ precesses in an effective magnetic field being the sum of the external and the Overhauser field with frequency ${\Omega_\mathrm{eff} \coloneqq | \bm\Omega_\mathrm{L} + \bm \Omega_\mathrm{N}|}$. The Larmor frequency $\bm\Omega_\mathrm{L}$ points along the external magnetic field. But the Overhauser field $\bm \Omega_\mathrm{N}$, whose time evolution is much slower than the pulse repetition rate~\cite{Merkulov2002}, has a random direction, which tilts the effective field from the $z$~axis due to its transverse components.
As illustrated by the sketch in Fig.~\ref{fig3}, this tilt leads to a precession motion, which becomes smaller in amplitude for a larger magnetic field. This is the reason why the higher RSA modes in Fig.~\ref{fig2} are less pronounced.
Strong pulses result in a strong generation of spin polarization while also aligning the electron spin along the $z$~axis, leading to RSA whenever the PSC $\Omega_\mathrm{eff} = |k| \wR$ for a single QD is met.
After averaging over the Overhauser field distribution described by Gaussian fluctuations with variance $\wn^2/2$~\cite{Merkulov2002,Stanek2014b}, the PSC~\eqref{PSC} follows in leading order for $\Omega_\mathrm{L} \gtrsim \wn$~\cite{Schering2019}. Corrections to the resonance frequency are~$O(\wn^2/\Omega_\mathrm{L}^2)$. For $\Omega_\mathrm{L} < \wn$, the modes appear shifted.
We point out that this mechanism is expected to work also for single QDs because the statistical fluctuations of the Overhauser field in time can be described by the same distribution as for the ensemble~\cite{Merkulov2002,Stanek2014b}.

RSA in Faraday geometry should be detectable in a variety of semiconductor nanostructures. 
But the ensemble average smears out the RSA modes such that they cannot be observed unless certain conditions are met.
The prerequisites are: (i)~an efficient hyperfine coupling, (ii)~strong pump pulses, and (iii)~a laser repetition frequency~$\wR$ which on the one hand allows for RSA modes that are separated enough not to overlap significantly, but on the other hand fall into magnetic field ranges where the spin polarization is not yet saturated~\cite{Schering2019}. The conditions~(i) and~(ii) are typically fulfilled for singly-charged $n$-type (In,Ga)As/GaAs~QDs~\cite{Urbaszek2013}. 
For $p$-type QDs with strongly anisotropic hyperfine interaction~\cite{Testelin2009,PhysRevB.101.115302}, the effect is exptected to be much weaker~\cite{Schering2019}.
The condition (iii) is not trivial and potentially leads to a new regime of the spin dynamics. An estimate for its realization is the condition \mbox{$\wR \gtrsim \sqrt{2}\wn$} known from standard RSA~\cite{Yugova2012}. We implemented a laser source with repetition rate \mbox{$\wR/(2\pi) = 1\,$GHz} to reach this regime. 

To provide a quantitative basis for the above, we study the RSA visibility defined as~\cite{Smirnov2020b}
\begin{align}
	V \coloneqq \frac{E_\mathrm{max} - E_\mathrm{min}}{E_\mathrm{max}} \,,
	\label{eq:visibility}
\end{align}
where $E_\mathrm{max}$ is the Faraday ellipticity of the first maximum at the RSA mode $|k|=1$ and $E_\mathrm{min}$ denotes the adjacent minimum for larger magnetic field $|\Bext|$.
In order to average out statistical fluctuations in both model and experiment, we fit polynomials to the region around the $|k| = 1$ mode to determine its visibility.
A map of the RSA visibility in dependence of the pulse area~$\Theta$ (defined in the Supplemental Material~\cite{Supplement}) and of the product~$\wn \TR$ ($\TR = 2\pi/\wR$ is the pulse repetition period) is shown in Fig.~\ref{fig3}. The visibility is enhanced by an increase of the pulse area corresponding to stronger pulses. For a too large or too small repetition period, the visibility decreases. Observing RSA in Faraday geometry is easiest in the intermediate regime $\wn \TR \sim 1 - 2$. The white cross on the color map shows that the laser source used here is operating at a too small power with a resulting visibility of merely $V \approx 0.02$ at a pump power of $P_\mathrm{pu} = 8.5\,$mW. Under optimal conditions, the visibility could reach unity. For the commonly used pulse repetition periods $\TR = 13.2$~and~$6.6\,$ns ($\wn\TR \approx 10.6$ and~$5.8$), no RSA modes are discernible for our QD sample with \mbox{$\wn/(2\pi) = 140\,$MHz} because they overlap.

\begin{figure}[t]
	\begin{center}
		\includegraphics[trim=0mm 0mm 0mm 0mm, clip, width=\columnwidth]{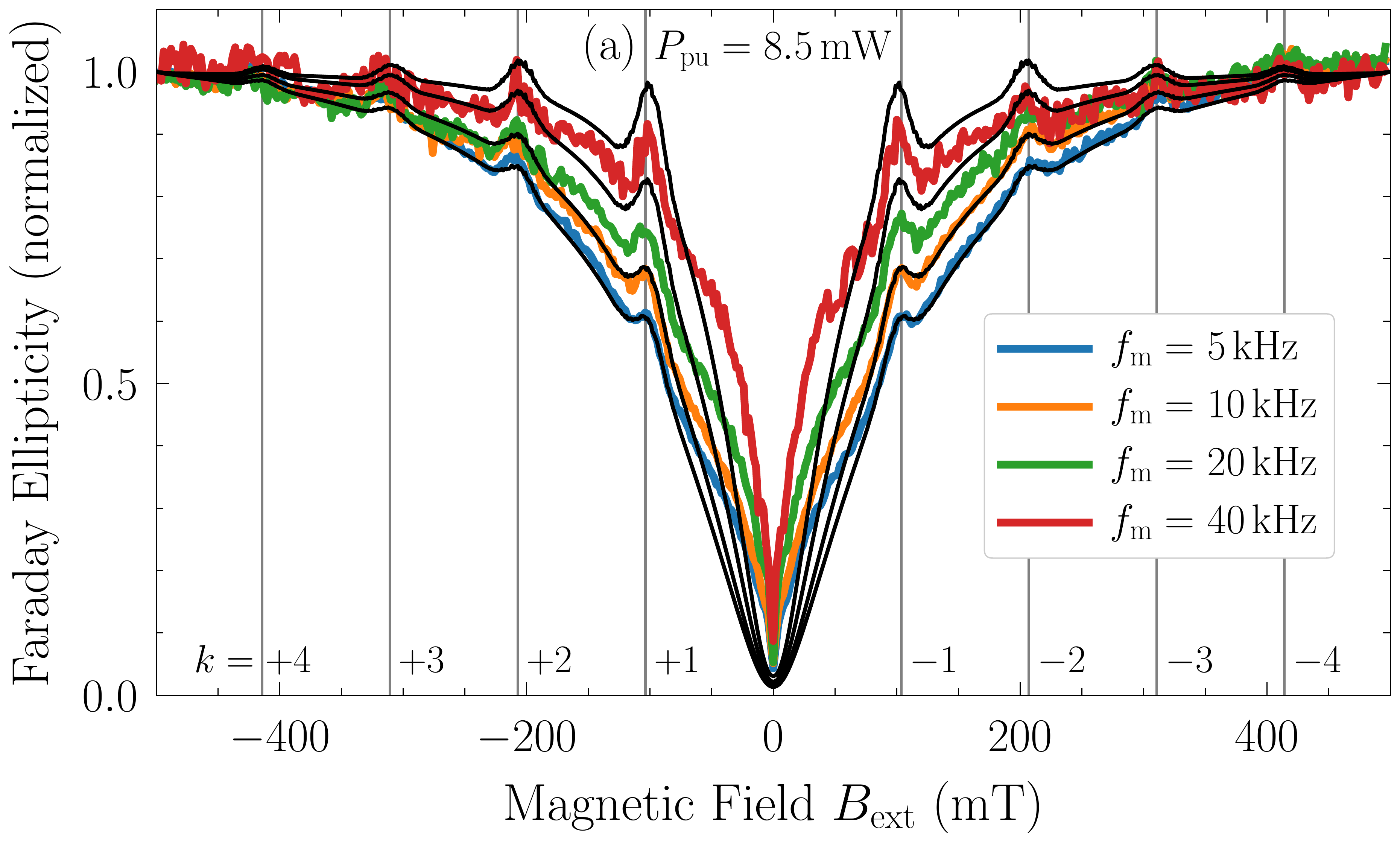}\\
		\includegraphics[trim=0mm 0mm 0mm 0mm, clip, width=\columnwidth]{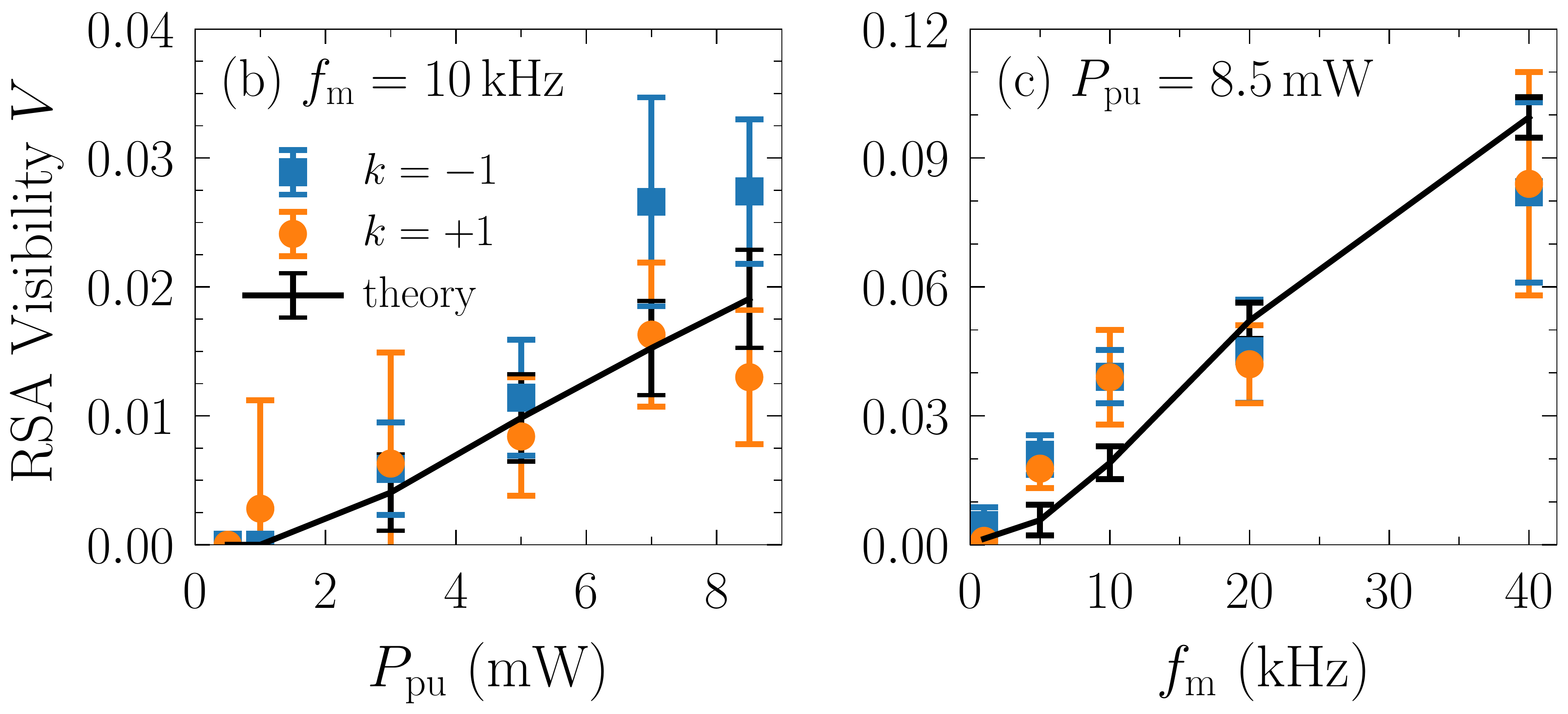}
		\caption{(a) Normalized PRCs for various pump modulation frequencies~$\fm$ at a pump power of $P_\mathrm{pu} = 8.5\,$mW. Except for the normalization with respect to the Faraday ellipticity at $\Bext = \pm 500\,$mT, the layout of the plot is analogous to Fig.~\ref{fig2}. Bottom row: RSA visibility $V$ as a function of the (b)~pump power~$P_\mathrm{pu}$ and (c)~pump modulation frequency~$\fm$ for the first RSA mode $|k| = 1$. The experimental data is plotted as blue squares ($k = -1$) and orange circles ($k = 1$), the theoretical data in black. The error bars represent the root-mean-square error of the polynomial fits around the respective RSA mode.}
		\label{fig4}
	\end{center}
\end{figure}

The main obstacle for a larger visibility is the small pulse area of the applied GHz~pulses because a too large pump power would heat the sample too much.
Yet, we can study the power dependence of the experimentally seen and theoretically modeled visibility in the accessible range.
Clearly, as demonstrated in Fig.~\ref{fig4}(b) by experiment and theory, a reduction of the pump power results in the disappearance of the RSA modes displayed by a vanishing visibility.

Remarkably, the RSA visibility can be enhanced by exploiting the spin-inertia effect. This is demonstrated in Fig.~\ref{fig4}(a) where PRCs for various pump modulation frequencies~$\fm$ are plotted using a pump power of $P_\mathrm{pu} = 8.5\,$mW. The PRCs are normalized with respect to the Faraday ellipticity at $\Bext = \pm 500\,$mT to highlight the enhanced visibility of the RSA modes upon increasing $\fm$. The deviation between experiment (colored) and theory (black) for large $\fm$ is related to the fact that the spin-inertia dependence for the QD sample is not perfectly captured by the monoexponential spin relaxation entering in our model.
But generally, the spin-inertia effect leads to a reduction of the average spin polarization upon increasing the modulation frequency~$\fm$ as shown earlier in Fig.~\ref{fig1}(b).
The key idea is the following: The application of a larger modulation frequency results in a decrease of the average absolute spin polarization and in turn, each pump pulse can better orient the spins along the optical axis because a larger number is disordered. In a nutshell, the spin-inertia effect allows us to avoid the influence of spin saturation, which is detrimental to RSA in Faraday geometry. Note also that for the same reason, a larger modulation frequency narrows the zero-field minimum of the PRCs as visible in Fig.~\ref{fig4}(a)~\cite{Schering2019}.
	
To be more quantitative, we plot the extracted RSA visibility~\eqref{eq:visibility} as a function of $\fm$ in Fig.~\ref{fig4}(c). Clearly, an increase of the modulation frequency results in a significant increase of the visibility~$V$, in agreement with the theoretical prediction. For instance, we find \mbox{$V = 0.08 \pm 0.02$} for the $k = +1$ mode using $\fm = 40\,$kHz.
The experimental limitation, especially for the higher RSA modes~$|k|$, is the deteriorated signal-to-noise ratio in the PRCs for larger frequencies as noticeable in Fig.~\ref{fig4}(a). The signal-to-noise ratio can be improved by averaging over longer time intervals for each data point.

Complementary results are provided in the Supplemental Material~\cite{Supplement}, which demonstrate RSA in Faraday geometry for another QD ensemble with a weaker nuclear fluctuation field. There, we study the Faraday rotation instead of the Faraday ellipticity, but we emphasize that the effect is easier to detect by measuring the ellipticity.

The PSC~\eqref{PSC} is also well known from the interrelated effects ``spin mode locking'' and ``nuclei-induced frequency focusing''~\cite{Greilich2006,Greilich2007,Greilich2009,Yugova2009,Barnes2011,Yugova2012,Glazov2012,Economou2014,Varwig2014,Beugeling2016,Beugeling2017,Schering2018,Kleinjohann2018,Schering2020,Vezvaee2020,Schering2021,Evers2021}. They occur in similar experimental setups but with the magnetic field being applied in Voigt geometry.
Since RSA takes place on much shorter time scales than the adaption of the Overhauser field responsible for nuclei-induced frequency focusing, we expect that it plays no role in our experiments. Yet, it is the subject of future research, which requires the inclusion of the nuclear spin dynamics in the model.

\paragraph*{Conclusion --}
\label{sec:conclusion}

We demonstrated that the detrimental nuclear spin fluctuations in QDs can be exploited for one's own advantage: they enable RSA in Faraday geometry.
The positions of the RSA modes directly yield the longitudinal $g$~factor of the charge carriers, which we determine very precisely to be $g_z = -0.69 \pm 0.01$ for the studied $n$-doped (In,Ga)As/GaAs QDs. 
This method solves the long-standing problem of measuring the longitudinal $g$~factor of the charge carriers directly in weak magnetic fields~\footnote{The effective $g$~factor may change due to band mixing, which is particularly relevant at high magnetic fields.} and puts the characterization of the spin dynamics in a longitudinal magnetic field on equal footing with the case of a transverse field.
The theoretical analysis paves the way to achieve a better visibility of the effect: use of strong pump pulses combined with a sufficiently high laser repetition rate.
A significant enhancement is achieved by exploiting the spin-inertia effect.
We believe that this technique will also be useful for the investigation of other semiconductor nanostructures, e.g., quantum wells.

\begin{acknowledgements}
We thank M.~M.~Glazov for helpful discussions and acknowledge the supply of the QD samples by D.~Reuter and A.~D.~Wieck.
P.S. and G.S.U. gratefully acknowledge the resources provided by the Gauss Centre for Supercomputing~e.\,V. on the supercomputer HAWK at High-Performance Computing Center Stuttgart and by the TU~Dortmund University on the HPC cluster LiDO3, partially funded by the Deutsche Forschungsgemeinschaft~(DFG) in Project No. 271512359. D.S.S. gratefully acknowledges the RF President Grant No. MK-5158.2021.1.2 and the Foundation for the Advancement of Theoretical Physics and Mathematics ``BASIS''. This work has been supported by the DFG in the frame of the International Collaborative Research Centre TRR 160 (Projects A1, A4, A7) and by the Russian Foundation for Basic Research (Grants Nos. 19-52-12038, 20-32-70048).
\end{acknowledgements}


%

\end{document}


\makeatletter
\renewcommand{\fnum@figure}{Supplementary~\figurename~\thefigure}
\makeatother	
\renewcommand{\theequation}{S\arabic{equation}}

\title{Supplemental Material: Resonant spin amplification in Faraday geometry}

\author{P.~Schering}
\affiliation{Condensed Matter Theory, Technische Universit\"at Dortmund, 44221 Dortmund, Germany}

\author{E.~Evers}
\affiliation{Experimentelle Physik 2, Technische Universit\"at Dortmund, 44221 Dortmund, Germany}

\author{V.~Nedelea}
\affiliation{Experimentelle Physik 2, Technische Universit\"at Dortmund, 44221 Dortmund, Germany}

\author{D.~S.~Smirnov}
\affiliation{Ioffe Institute, Russian Academy of Sciences, 194021 St.\,Petersburg, Russia}

\author{E.~A.~Zhukov}
\affiliation{Experimentelle Physik 2, Technische Universit\"at Dortmund, 44221 Dortmund, Germany}
\affiliation{Ioffe Institute, Russian Academy of Sciences, 194021 St.\,Petersburg, Russia}

\author{D.~R.~Yakovlev}
\affiliation{Experimentelle Physik 2, Technische Universit\"at Dortmund, 44221 Dortmund, Germany}
\affiliation{Ioffe Institute, Russian Academy of Sciences, 194021 St.\,Petersburg, Russia}

\author{M.~Bayer}
\affiliation{Experimentelle Physik 2, Technische Universit\"at Dortmund, 44221 Dortmund, Germany}
\affiliation{Ioffe Institute, Russian Academy of Sciences, 194021 St.\,Petersburg, Russia}

\author{G.~S.~Uhrig}
\affiliation{Condensed Matter Theory, Technische Universit\"at Dortmund, 44221 Dortmund, Germany}

\author{A.~Greilich}
\affiliation{Experimentelle Physik 2, Technische Universit\"at Dortmund, 44221 Dortmund, Germany}

\date{\today}

\begin{abstract}
\end{abstract}

\maketitle

The Supplemental Material is divided into two parts. In Sec.~\ref{sec:model}, we provide all the details of the model used to analyze the performed experiments which reveal resonant spin amplification~(RSA) in Faraday geometry. Complementary results, which demonstrate RSA in Faraday geometry detected by Faraday rotation for another quantum dot~(QD) ensemble, are presented in Sec.~\ref{sec:complementary_results}. We will refer to this QD ensemble as `sample~B' and to the one studied in the main text as `sample~A'.

\section{Theoretical model}
\label{sec:model}

The applied theoretical model is a combination of the model for the spin dynamics used in Ref.~\cite{Schering2019} with the formalism of Ref.~\cite{Yugova2009} which allows for a detailed description of the pumping and probing of spins in QD ensembles. As a novel ingredient, we account for trion states with a lifetime comparable to the pulse repetition period.
The layout of this section is as follows. 
First, we introduce the equations of motion describing the spin dynamics for a single QD which is singly charged by an electron.
Then, the pulse model to describe the optical spin orientation induced by a pump pulse is introduced.
In order to model the experimental setup, we repeat the formalism of Ref.~\cite{Yugova2009} which allows us to calculate the experimentally probed Faraday ellipticity and rotation for a single QD.
Subsequently, averaging over the inhomogeneous ensemble of QDs takes place.
Finally, we have to account for the experimental detail that an accumulated signal is measured and the helicity of the pump pulses is modulated.

\subsection{Statement of the theoretical problem}

We consider an ensemble of (In,Ga)As/GaAs QDs singly charged by electrons ($n$~doped) in a pump-probe type experiment as described in the main text and in Sec.~\ref{sec:complementary_results}.
A magnetic field is applied along the axis of light propagation ($z$~axis, Faraday geometry).
The periodic pump pulses excite singlet trion states leading to the spin orientation of the resident electrons according to the optical selection rules~\cite{Ivchenko2005a}. 
The spin polarization along the axis of light propagation is probed by weak linearly polarized pulses measuring the Faraday ellipticity or rotation~\cite{Yugova2009,Glazov2012b}.
The difference between these two quantities is that they probe the polarization of different subsets of the QD ensemble, depending on the detuning between the pump and probe pulses~\cite{Yugova2009}.
In our case, the pump and probe pulses have the same photon energy.

The localized spins in the QDs are excited by very long trains of pump pulses following one another with repetition period $\TR$. 
Each pulse is circularly polarized and the helicity of the pulses is alternated with modulation frequency~$\fm$.
The spin polarization is probed by weak probe pulses with the same repetition period.
Experimentally, they arrive shortly before the next pump pulse with a delay of $-60\,$ps (main text) or $-50\,$ps~(Sec.~\ref{sec:complementary_results}). 

\subsection{Equations of motion for the spin dynamics in a single quantum dot}

We apply the same phenomenological model for the spin dynamics as used in Ref.~\cite{Schering2019}.
Here, we consider only electrons as resident charge carriers ($n$ doped); see Ref.~\cite{Schering2019} for the model describing $p$-doped QDs.
Our main focus is on the electron spin dynamics, but the photoexcited trion determines the spin generation rate, which is essential for the physics~\cite{Smirnov2018,Zhukov2018}.

The dynamics between consecutive pump pulses of a single electron spin $\S$ in a QD can be described by the equation of motion
\begin{equation}
\label{eq:dSdt}
\frac{d}{d t} \S = \left(\bm\Omega_\mathrm{N}+\bm\Omega_\mathrm{L}\right)\times\S 
- \frac{\S}{\tausg} + \frac{J^z}{\tau_0} \ez \,,
\end{equation}
where $\bm\Omega_\mathrm{N}$ is the frequency of the spin precession in the random nuclear field (Overhauser field) due to the hyperfine interaction~(HI) and ${\bm\Omega_\mathrm{L}= \Omega_\mathrm{L}\ez = g_z \mu_\mathrm{B} \Bext \hbar^{-1} \ez}$ 
is the Larmor frequency due to the external longitudinal magnetic field~$\Bext$ ($g_z$~is the longitudinal electronic $g$~factor,
$\mu_\mathrm{B}$~the Bohr magneton, $\hbar$~the reduced Planck constant).
Furthermore, the phenomenological term $-\S/\tausg$ describes the spin relaxation
unrelated to the hyperfine interaction with the nuclear spins in the QD. 
The component~$J^z$ is the $z$~projection of the trion pseudospin~$\bm J$ and $\tau_0 = 0.4\,$ns~\cite{greilichPRB06,Greilich2006a} is the radiative trion lifetime.
The last term in Eq.~\eqref{eq:dSdt} describes the input of spin polarization due to the radiative trion recombination~\cite{Yugova2009,Zhukov2007}. 
We neglect the explicit dynamics of the Overhauser field and assume it to be static~\cite{Merkulov2002}.

The pump pulses optically excite the electron leading to the formation of a negatively charged trion~$\mathrm{T}^-$. It consists of two electrons in a spin singlet and a heavy hole with unpaired spin, which can be represented by an effective pseudospin $\J$~\cite{Glazov2012b}.
Its dynamics between consecutive pump pulses is described similarly to Eq.~\eqref{eq:dSdt} by
\begin{equation}
\label{eq:dJdt}
\frac{d}{d t}\J =\left(\bm\Omega_\mathrm{N}^\mathrm{T}+\bm\Omega_\mathrm{L}^\mathrm{T}\right)\times\J - \frac{\bm J}{\taust} - \frac{\bm J}{\tau_0} \,.
\end{equation}
The label `T' refers to the parameters of the heavy hole spin in the trion.
The nonradiative trion recombination, which does not contribute to the spin polarization in the ground state and which is unrelated to the hyperfine interaction, is accounted for by the relaxation time $\taust$. 
Combined with Eq.~\eqref{eq:dSdt} and according to the optical selection rules, only the $z$~component of the trion pseudospin~$\J$ is transferred back to the ground state $\S$ on the timescale $\tau_0$ due to the radiative trion decay.

Due to the large number of $O(10^5)$ nuclear spins in each QD coupled to the electron spin via the hyperfine interaction, the Overhauser field can be treated as a classical fluctuation field~\cite{Merkulov2002,CSM}. According to the central limit theorem, its probability distribution is Gaussian~\cite{Schulten1976,Merkulov2002,Stanek2014b},
\begin{align}
p(\bm\Omega_\mathrm{N}) = \frac{\lambda^2}{(\sqrt{\pi} \omega_\mathrm{N})^3} \exp\left(-\lambda^2\frac{(\Omega_\mathrm{N}^x)^2+(\Omega_\mathrm{N}^y)^2}{\omega_\mathrm{N}^2} - \frac{(\Omega_\mathrm{N}^z)^2}{\omega_\mathrm{N}^2}\right) \,,
\label{eq:Overhauser_distribution}
\end{align}
where $\wn^2/2$ is the variance and $\lambda$ parameterizes the potential degree of anisotropy of the hyperfine interaction. For electrons the hyperfine interaction is isotropic, i.e., $\lambda=1$.
Generally, the frequency $\wn$ represents the characteristic fluctuation strength of the nuclear spin bath seen by the electron spin.
In contrast, for heavy holes the hyperfine interaction is strongly anisotropic with $\lambda>1$~\cite{Testelin2009,PhysRevB.101.115302}.
Thus, since the trion spin precession is related to the same nuclear spin bath, it can be assumed that it is given by
\begin{subequations}
	\begin{align}
	\Omega_\mathrm{N}^{\mathrm{T},x} &= \chi\frac{\lambda}{\lambt}\Omega_\mathrm{N}^x \,,
	\qquad 
	\Omega_\mathrm{N}^{\mathrm{T},y} = \chi\frac{\lambda}{\lambt}\Omega_\mathrm{N}^y \,,\\
	\Omega_\mathrm{N}^{\mathrm{T},z} &= \chi\Omega_\mathrm{N}^z \,,
	\end{align}
\end{subequations}
where $\chi=\wt/\wn$ describes the relative strength of the hyperfine interaction for the trion state and $\lambda^\mathrm{T}$ quantifies its anisotropy.
Physically, the averaging over the distribution~\eqref{eq:Overhauser_distribution} models two cases: (i) the average over a homogeneous QD ensemble and (ii) the time average in an experiment where the signal is probed over a time much longer than the typical correlation time of the nuclei (about $200\,$ns for sample~B~\cite{Zhukov2018}).

For $n$-doped QDs, it is typically sufficient to neglect the precession-term in Eq.~\eqref{eq:dJdt}.
The reason is that in this case, the trion pseudospin stems from the heavy hole with weak and anisotropic hyperfine interaction, i.e., its contribution to the generation rate of spin polarization is negligible~\cite{Smirnov2018}.
The ensemble of (In,Ga)As/GaAs QDs studied in Sec.~\ref{sec:complementary_results} (sample~B) is well characterized~\cite{Zhukov2018}; this is not the case for sample~A studied in the main text so that more parameters need to be fitted.
Hence, we neglect the precession-term in Eq.~\eqref{eq:dJdt} in the analysis of sample~A but include it for sample~B. 
Despite the better characterization of sample~B, the results for sample~A shown in the main text are more suitable to demonstrate the effect of RSA in Faraday geometry due to a stronger hyperfine interaction leading to more visible RSA modes.

In our model calculations presented in Sec.~\ref{sec:complementary_results}, we apply the system parameters listed in Table~\ref{tab:parameters}.
The two fit parameters varied to reproduce the experimental results are the longitudinal electronic $g$~factor, which is determined by the positions of the experimentally observed RSA modes, and the trion spin relaxation time $\taust$ (estimated in Ref.~\cite{Zhukov2018} to be $\taust < 1000\,$ns).

For sample~A studied in the main text, the longitudinal electronic $g$~factor is also determined by the positions of the RSA modes.
The parameter~$\wn$ is estimated based on an extrapolation of the polarization recovery curve~(PRC) to zero pump power; the spin relaxation time~$\tausg$ is determined from spin-inertia measurements~\cite{Schering2019}. The applied parameters are also summarized in Table~\ref{tab:parameters}.

We point out that the simulated PRCs are fairly sensitive to the choice of~$\taust$ and of the effective pulse area~$\Theta$ (defined in the following), and both quantities can only be estimated. It is possible that other combinations exist which yield a similarly good agreement between simulation and experiment.

\begin{table}[t]
	\caption{System parameters and their physical meaning used in the model calculations for the two different $n$-doped (In,Ga)As/GaAs QD ensembles studied in the main text and in Sec.~\ref{sec:complementary_results}, respectively. The parameters used in Sec.~\ref{sec:complementary_results} are based on a previous sample characterization~\cite{Zhukov2018}.}
	\label{tab:parameters}
	\begin{ruledtabular}
	\begin{tabular}{lrl}
		\textbf{Parameter} & \textbf{Value} & \textbf{Physical meaning}\\
		\midrule
		\multicolumn{3}{l}{\textbf{Sample A (main text)}} \\
		$\wn/(2\pi)$ & $140\,\mathrm{MHz}$ & electron HI strength\\
		$\tausg$ & $22\,\mu\mathrm{s}$ & electron spin relaxation time\\
		$\taust$ & $0.45\,\mu\mathrm{s}$ & hole spin relaxation time\\
		$g_z$ & $-0.69$ & longitudinal electron $g$~factor\\
		\midrule
		\multicolumn{3}{l}{\textbf{Sample B (Sec.~\ref{sec:complementary_results})}} \\
		$\wn/(2\pi)$ & $70\,\mathrm{MHz}$ & electron HI strength\\
		$\wt/(2\pi)$ & 1$6\,\mathrm{MHz}$ & hole HI strength \\
		$\lambt$ & $5$ & hole HI anisotropy\\ 
		$\tausg$ & $1.3\,\mu\mathrm{s}$ & electron spin relaxation time\\
		$\taust$ & $0.06\,\mu\mathrm{s}$ & hole spin relaxation time\\
		$g_z$ & $-0.64$ & longitudinal electron $g$~factor\\
		$g_z^\mathrm{T}$ & $-0.45$ & longitudinal hole $g$~factor\\		
		\midrule
		\multicolumn{3}{l}{\textbf{Common parameters}} \\
		$\tau_0$ & $0.4\,\mathrm{ns}$ & radiative trion lifetime\\	
		$\lambg$ & $1$ & electron HI anisotropy\\		
	\end{tabular}	
	\end{ruledtabular}
\end{table}

\subsection{Spin polarization induced by optical trion excitation}

Next, we turn to the description of the action of a circularly polarized pump pulse on the localized electron spin in a single QD.
In the experiments, a laser with a pulse repetition period of $\TR = 1\,$ns is used.
In this case, the optically excited trion state is still slightly populated before the arrival of the next pulse due to its comparable lifetime $\tau_0 = 0.4\,$ns.
Hence, we cannot straightforwardly apply the pulse model of Ref.~\cite{Yugova2009}. Instead, we generalize it to account for a finite trion population at the arrival of a pump pulse. The generalized pulse relations are introduced in the following.

To this end, we introduce the occupation numbers of the ground and excited trion state, whose recombination dynamics between consecutive pulses are simply described by
\begin{subequations}
	\begin{align}
	n^\mathrm{G}(t) &= n_\mathrm{a}^\mathrm{G} + n_\mathrm{a}^\mathrm{T} \left[1 - \exp\left(-\frac{t}{\tau_0}\right) \right] \,, \\
	n^\mathrm{T}(t) &= n_\mathrm{a}^\mathrm{T} \exp\left(-\frac{t}{\tau_0}\right) \,.
	\end{align}
\end{subequations}
Here, $n_\mathrm{a}^\mathrm{G}$ and $n_\mathrm{a}^\mathrm{T}$ denote the occupation numbers of the ground and trion state immediately after (label~`a') a pump pulse.
Before the arrival of the very first pulse, we start from the initial conditions $\S_\mathrm{b} = 0$, $\J_\mathrm{b} = 0$, $n_\mathrm{b}^\mathrm{G} = 1$, and $n_\mathrm{b}^\mathrm{T} = 0$ in our model calculations (the label~`b' denotes the value before a pump pulse), i.e., there is no finite polarization and all electron spins are in their ground state. The initial condition represents a completely disordered state.

The pulses in the experiment have a duration of $2\,$ps, which is much shorter than all other timescales of the system. Hence, we can treat them as instantaneous pulses.
Under this condition and following the formalism of Ref.~\cite{Yugova2009}, the spin components of the electron spin before~($\S_\mathrm{b}$,~$\J_\mathrm{b}$) and after~($\S_\mathrm{a}$,~$\J_\mathrm{a}$) a pump pulse can be related to each other by
\begin{subequations}
	\begin{align}
	S^x_\mathrm{a} &= Q \cos(\Phi) S^x_\mathrm{b} + \mathcal{P} Q \sin(\Phi) S^y_\mathrm{b} \,,\\
	S^y_\mathrm{a} &= Q \cos(\Phi) S^y_\mathrm{b} - \mathcal{P} Q \sin(\Phi) S^x_\mathrm{b} \,,\\
	S^z_\mathrm{a} &= - \mathcal{P} \frac{1 - Q^2}{4} (n_\mathrm{b}^\mathrm{G} - n_\mathrm{b}^\mathrm{T}) + \frac{1+Q^2}{2} S^z_\mathrm{b}	+ \frac{1-Q^2}{2} J^z_\mathrm{b} \,,
	\end{align}
\end{subequations}
and similary for the trion pseudospin by
\begin{subequations}
	\begin{align}
	J^x_\mathrm{a} &= Q \cos(\Phi) J^x_\mathrm{b} - \mathcal{P} Q \sin(\Phi) J^y_\mathrm{b} \,,\\
	J^y_\mathrm{a} &= Q \cos(\Phi) J^y_\mathrm{b} + \mathcal{P} Q \sin(\Phi) J^x_\mathrm{b} \,,\\
	J^z_\mathrm{a} &= \mathcal{P} \frac{1-Q^2}{4} (n_\mathrm{b}^\mathrm{G} - n_\mathrm{b}^\mathrm{T}) + \frac{Q^2+1}{2} J^z_\mathrm{b} + \frac{1-Q^2}{2} S^z_\mathrm{b} \,.
	\end{align}
\end{subequations}
These relations depend on the occupation numbers of the ground and trion state, which are also affected by the pump pulse. Their values before~($n_\mathrm{b}^\mathrm{G}$,~$n_\mathrm{b}^\mathrm{T}$) and after~($n_\mathrm{a}^\mathrm{G}$,~$n_\mathrm{a}^\mathrm{T}$) the pulse are related by
\begin{subequations}
	\begin{align}
	n_\mathrm{a}^\mathrm{G} &= \frac{1+Q^2}{2} n_\mathrm{b}^\mathrm{G} + \frac{1-Q^2}{2} n_\mathrm{b}^\mathrm{T} - \mathcal{P} (1 - Q^2) (S^z_\mathrm{b} - J^z_\mathrm{b}) \,, \\
	n_\mathrm{a}^\mathrm{T} &= \frac{1+Q^2}{2} n_\mathrm{b}^\mathrm{T} + \frac{1-Q^2}{2} n_\mathrm{b}^\mathrm{G} + \mathcal{P} (1 - Q^2) (S^z_\mathrm{b} - J^z_\mathrm{b}) \,.
	\end{align}
\end{subequations}
Here, $0\le Q^2 \le 1$ describes the probability for the pulse not to excite a trion, i.e., $Q$ is related to the pumping efficiency. For $Q=0$, which is the most efficient case, each pulse completely aligns the electron spin $\S$ along the $z$~axis because the transverse components are set to zero.
The parameter $\Phi$ describes the spin rotation induced by detuned pulses, and $\mathcal{P}=\pm1$ represents the helicity of the circularly polarized pulse. 
The pulse parameters $Q$ and $\Phi$ depend on the detuning of the pump pulse from the trion transition energy. This is discussed in the following subsection.
Note that one easily retains the pulse relations of Ref.~\cite{Yugova2009} by inserting $n_\mathrm{b}^\mathrm{G} = 1$, $n_\mathrm{b}^\mathrm{T} = 0$, and $\J_\mathrm{b} = 0$. This simplification would be valid for $\tau_0 \ll \TR$, e.g., for the commonly used pulse repetition period~$\TR = 13.2\,$ns. However, this is not the case for the experiments described in this work with $\TR = 1\,$ns.

\subsection{Inhomogeneous ensemble of quantum dots}

All QDs in a real ensemble are slightly different, e.g., in size, shape, composition, and strain.
In particular, due to the inhomogeneous broadening of the trion transition, each QD of the ensemble has a slightly different trion transition energy~$E_\mathrm{T}$.
We model this situation by assuming that the transition energies are normally distributed according to
\begin{align}
\label{eq:trion_distribution}
p(E_\mathrm{T}) = \frac{1}{\sqrt{ 2\pi \sigma^2_\mathrm{E_\mathrm{T}} }} \exp\left(-\frac{(E_\mathrm{T} - \mu_\mathrm{E_\mathrm{T}})^2}{2 \sigma^2_\mathrm{E_\mathrm{T}}}\right) \,.
\end{align}
This situation implies that for a fixed energy of the pump pulses all the individual QDs are pumped with different efficiencies due to detuning between laser and trion transition energy. Eventually, this can be described by an associated pair~$\{Q,\,\Phi\}$ of the pulse parameters for each QD.
Similarly for the probe pulses, different QDs have a different contribution to the measured Faraday ellipticity and rotation. Tuning the pump and probe energies allows for an analysis of different subsets of the QD ensemble.
For precise details, we refer the interested reader to Ref.~\cite{Yugova2009} where this formalism is established.

\subsubsection{Pulse parameters for a single quantum dot}

The energies of the pump and probe pulse are denoted as $E_\mathrm{pu}$ and $E_\mathrm{pr}$, respectively.
In the experiments, a degenerate pump-probe setup is used, i.e., $E_\mathrm{pu} = E_\mathrm{pr}$.
The finite spectral width of the pulses is accounted for by the inverse pulse duration $\tau_\mathrm{p}^{-1}$, i.e., we assume that the pulses are Fourier-transform limited.
For simplicity, we model the pulse shape as Rosen-Zener pulses~\cite{RosenZener} for which one can derive analytical expressions for the pulse parameters $Q$ and $\Phi$~\cite{Yugova2009}.
The amplitude of the electric field of such pulses with duration $\tau_\mathrm{p}$ has the form
\begin{align}
	f(t) = \mu \, \mathrm{sech}\left( \pi \frac{t}{\tau_\mathrm{p}} \right) \,, \label{eq:sech}
\end{align}
where $\mu$ is a measure for the electric field strength.
In the experiments, the actual pulse shape is Gaussian and the pulses are not perfectly Fourier-transform limited.
We use $\tau_\mathrm{p} = 1.3\,$ps to account for the finite spectral width of the pulses.
For the pulse shape~\eqref{eq:sech}, it follows for the pulse parameters~\cite{Yugova2009}
\begin{subequations}
\begin{align}
	Q &= \sqrt{1 - \frac{\sin^2(\Theta/2)}{\cosh^2(\pi y)}} \,, \\
	\Phi &= \arg \left[ \frac{\Gamma^2\left(\frac{1}{2} - i y\right)}{\Gamma\left(\frac{1}{2} - \frac{\Theta}{2\pi} - i y\right) \Gamma\left(\frac{1}{2} + \frac{\Theta}{2\pi} - i y\right)} \right] \,,
\end{align}
\end{subequations}
where $\Gamma(z)$ is the Gamma function, 
\begin{align}
	\Theta = 2 \int_{-\infty}^\infty f(t') dt' = 2 \mu \tau_\mathrm{p}
\end{align}
is the effective pulse area, and
\begin{align}
	y = \frac{(E_\mathrm{pu} - E_\mathrm{T}) \tau_\mathrm{p}}{2\pi \hbar}
\end{align}
is the dimensionless detuning of the pump pulses.

For small pump powers $P_\mathrm{pu}$, the pulse area scales like ${\Theta \propto \sqrt{P_\mathrm{pu}}}$~\cite{Yugova2009}. 
We use this scaling to obtain a first estimate of the pulse area in the experiments.
For the experiments described in the main text, the best fit is achieved using $\Theta = 0.07\pi$ for a pump power of $P_\mathrm{pu} = 8.5\,$mW.
For the experiments described in Sec.~\ref{sec:complementary_results}, we use $\Theta \approx 0.18\pi$ for a pump power of $P_\mathrm{pu} = 5\,$mW. The pulse areas for the other pump powers are chosen such that the resulting spin polarization for large magnetic fields fits the pump power dependence observed in experiment.

\subsubsection{Probing the Faraday ellipticity and Faraday rotation}

The Faraday ellipticity and rotation, which are proportional to the spin polarization, can be probed by weak linearly polarized pulses in the experiments.
Depending on the energy $E_\mathrm{pr}$ of the probe pulse, different subsets of the QD ensemble are probed, i.e., the contribution of a single QD to the probed signal depends on the detuning of the probe pulse from the trion transition energy~$E_\mathrm{T}$.
In the experiments, the energy of pump and probe are degenerate, i.e., $E_\mathrm{pu} = E_\mathrm{pr}$.

For a single QD, the Faraday ellipticity $\E$ and rotation $\R$ are proportional to the spin polarization $J^z - S^z$ weighted by an additional function which depends on the probe detuning and pulse duration~\cite{Yugova2009},
\begin{subequations}
\begin{align}
\E &\propto (J^z - S^z) \,\mathrm{Re}\, G(E_\mathrm{pr} - E_\mathrm{T}, \tau_\mathrm{p}) \label{eq:E} \,,\\
\R &\propto (J^z - S^z) \,\mathrm{Im}\,G(E_\mathrm{pr} - E_\mathrm{T}, \tau_\mathrm{p}) \label{eq:F} \,, 
\end{align}
\label{eq:FE}
\end{subequations}
with 
\begin{align}
	G(E_\mathrm{pr} - E_\mathrm{T}, \tau_\mathrm{p}) = \frac{\tau_\mathrm{p}^2}{\pi^2} \,\zeta\left(2,\, \frac{1}{2} - i \frac{(E_\mathrm{pr} - E_\mathrm{T}) \tau_\mathrm{p}}{2 \pi \hbar}\right) \,,
\end{align}
where $\zeta(z)$ is the Hurwitz Zeta function. 
With respect to the detuning $E_\mathrm{pr} - E_\mathrm{T}$, the real part \mbox{$\mathrm{Re}\,G(E_\mathrm{pr} - E_\mathrm{T}, \tau_\mathrm{p})$} is symmetric around zero while the imaginary part \mbox{$\mathrm{Im}\,G(E_\mathrm{pr} - E_\mathrm{T}, \tau_\mathrm{p})$} is antisymmetric.
The prefactors in Eq.~\eqref{eq:FE} are identical but sample dependent. 
The exact prefactors do not matter for our considerations because we scale the simulated polarization recovery curves to the experimental data by a global factor.

\subsubsection{Ensemble average}

In order to correctly account for the ensemble of QDs, two averaging processes must take place.
First, we have to average over the fluctuations of the nuclear spin bath described by the distribution~\eqref{eq:Overhauser_distribution}.
Second, we have to average over the distribution of the trion transition energies~\eqref{eq:trion_distribution}. For this purpose, we calculate the ensemble average over $2.6 \times 10^5$ (sample~A) or $10^6$ (sample~B) independent trajectories starting from random initial conditions sampled from these two distributions.
Numerically, this is extremely expensive and requires a massive parallelization.
The parameters $E_\mathrm{T}$, $\sigma_\mathrm{E_\mathrm{T}}$, $E_\mathrm{pu}$, $E_\mathrm{pr}$, and $\tau_\mathrm{p}$ are obtained by fitting the assumed shapes to the measured photoluminescence spectra.

As mentioned before, the Faraday ellipticity and rotation reveal the spin dynamics of different subsets of the QD ensemble~\cite{Yugova2009}.
In the experiments presented in Sec.~\ref{sec:complementary_results}, a degenerate pump-probe setup is used with the pulse energy being shifted by $2.4\,$meV to the low-energy flank of the average trion transition energy of about $1.3908\,$eV. 
In this case, the measured Faraday rotation effectively reveals the dynamics of the QDs which are excited by detuned pulses, i.e., the spin pumping efficiency is rather weak~\cite{Yugova2009}. However, as demonstrated in Ref.~\cite{Schering2019}, the spin rotation $\Phi$ induced by detuned pulses increases the visibility of RSA in Faraday geometry.
In contrast, by measuring the Faraday ellipticity as in the main text, the applied degenerate pump-probe scheme yields the spin polarization of a QD sub-ensemble which is pumped with almost zero detuning on average so that the pumping efficiency~$Q$ is enhanced (given the pulse area $\Theta$ is identical).

For the degenerate pump-probe setup, the Faraday ellipticity yields a better RSA visibility than the Faraday rotation.
It is the subject of future research to determine if a controlled pump-probe detuning~\cite{Glazov2010} can be used to enhance the visibility.

\subsection{Accumulated signal in the modulated pulse scheme}

In the experiments, the helicity of the pump pulses is modulated with the frequency~$\fm$ to enhance RSA and prevent the build-up of dynamic nuclear polarization. This modulation scheme leads to the formation of alternating spin polarization, which is zero on average~\cite{Smirnov2018}. For this reason, the probed signal must also be modulated. 
The actually measured Faraday ellipticity~$E$ or rotation~$R$ is given by~\cite{Heisterkamp2015,Smirnov2018}
\begin{subequations}
\begin{align}
	E(\fm) &= \frac{1}{n_\mathrm{p}} \left| \sum_{k=1}^{n_\mathrm{p}} \overline{\E(k\TR + \tau_\mathrm{d})} 
	e^{i 2\pi \fm (k \TR + \tau_\mathrm{d})} \right| \,, \\
	R(\fm) &= \frac{1}{n_\mathrm{p}} \left| \sum_{k=1}^{n_\mathrm{p}} \overline{\mathcal{R}(k\TR + \tau_\mathrm{d})} 
	e^{i 2\pi \fm (k \TR + \tau_\mathrm{d})} \right| \,.
	\label{eq:FR_def}
\end{align}
\end{subequations}
These expressions represent the accumulation of the probed signal for the ensemble of QDs (the ensemble average is denoted by the overline), modulated with frequency~$\fm$ and averaged over the number of applied pulses $n_\mathrm{p} \gg 1$. The Faraday rotation is probed slightly before the arrival of each pump pulse with a negative delay $\tau_\mathrm{d} = -60\,$ps (sample~A) or $-50\,$ps (sample~B).

In our numerical simulations, it is not feasible to calculate the spin dynamics for more than a few modulation periods $1/\fm$ while in the experiments, the pumping is performed over many modulation periods.
However, for small enough modulation frequencies as applied in experiment it is sufficient to simulate only two modulation periods and then to sum over the last period. The first period is neglected because it shows a slight transient behavior due to the simulations starting from equilibrium conditions.

\section{Complementary results}
\label{sec:complementary_results}

In this section, we complement the results of the main text by studying another (In,Ga)As/GaAs QD ensemble (sample~B) whose parameters are well known~\cite{Zhukov2018}. Resonant spin amplification in Faraday geometry is also demonstrated for this sample. In contrast to the experiments presented in the main text, we probe the Faraday rotation instead of the Faraday ellipticity to detect the spin polarization.

\subsection{Experimental details}
\label{ssec:experimental-details}
Sample~B consists of 20 layers of (In,Ga)As QDs separated by $60$-nm barriers of GaAs and grown by molecular beam epitaxy on a (100)-oriented GaAs substrate. A $\updelta$-doping layer of silicon $16\,$nm above each QD layer provides a single electron per QD on average. Rapid thermal annealing at $945\,^{\circ}$C for $30\,$s homogenizes the QD size distribution and shifts the average emission energy to $1.3908\,$eV. The QD density per layer amounts to~$10^{10}\,\text{cm}^{-2}$. 

\begin{figure}[b]
	\begin{center}
		\includegraphics[trim=0mm 0mm 0mm 0mm, clip, width=\columnwidth]{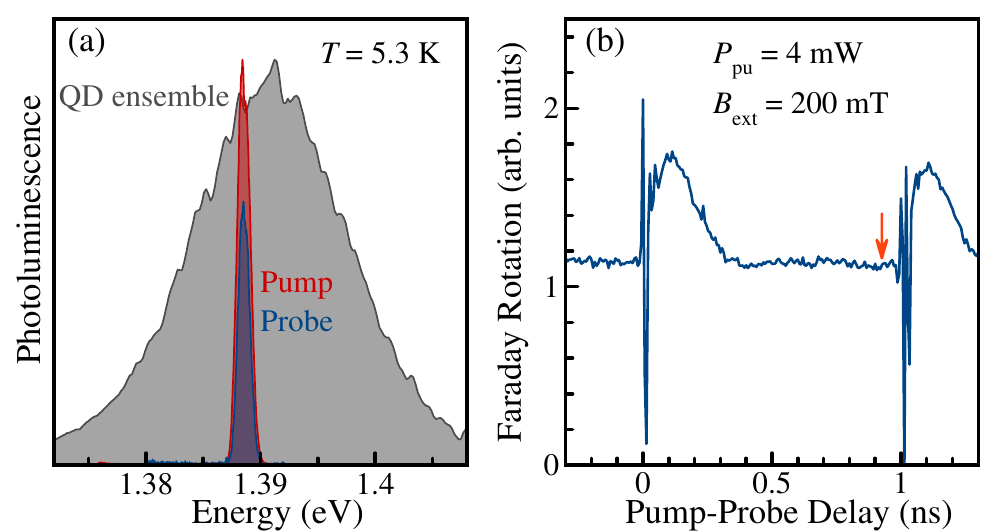}
		\caption{(a)~Photoluminescence of the QD ensemble (sample~B, gray) at \mbox{$T=5.3\,$K} along with the spectra for pump (red) and probe (blue) light. (b)~Time-resolved Faraday rotation with pump and probe pulse coincidence at 0 and $1\,$ns delay at a longitudinal magnetic field of \mbox{$\Bext=200\,$mT}. The orange arrow marks the delay ($-50\,$ps) at which the polarization recovery curves are taken.}
		\label{fig1}
	\end{center}
\end{figure}

The setup is very similar to the experimental details provided in the main text; the differences are given in the following.
The sample is illuminated with a central optical energy of $1.3884\,$eV and a full width at half maximum of $1.3\,$meV. 
The pump and the probe pulses with a duration of $2\,$ps are degenerate in photon energy and shifted by $2.4\,$meV to the low-energy flank of the QD photoluminescence, see Suppl.~Fig.~\ref{fig1}(a).
The helicity of the pump pulses is modulated by a photo-elastic modulator at $84\,$kHz between left- and right-handed circular polarization. 
The linearly polarized probe beam is modulated in intensity using an electro-optical modulator at a frequency of $50\,$kHz in series with a Glan prism and the reference frequency of the lock-in amplifier is running at the difference frequency of $34\,$kHz. The pump and the probe beams are focused on the same sample spot. The pump is focused to a spot diameter of $50\,\mu$m and the probe to a diameter of $45\,\mu$m. The Faraday rotation amplitude of the probe pulses is measured using an optical polarization bridge which consists of a Wollaston prism and a balanced photodetector.

We point out that our setup differs from the earlier experiments of Ref.~\cite{Zhukov2012} on CdTe quantum wells where the external magnetic field itself is tilted from the optical axis such that different components of the $g$~factor tensor determine the electronic Larmor frequency; in our case the external field points purely along the optical axis with an accuracy of 2 degree (Faraday geometry).

\subsection{Results}
\label{ssec:results}

As discussed in the main text, each circularly polarized pump pulse partially aligns the spin polarization of the resident electrons along the optical axis~\cite{Shabaev2003}.
The spin polarization added by every pulse in this way depends on its effective pulse area $\Theta$~\cite{Greilich2006a}, scaling like $P_{\text{pu}} \propto \Theta^2$ for small powers~\cite{Yugova2009}; see also Sec.~\ref{sec:model} and Fig.~\ref{fig3}(d).
For a train of pump pulses, the spin polarization builds up when the relaxed polarization along the optical axis between two pulses is smaller than the polarization added by a pump pulse. This is the case for an external field of $\Bext=200\,$mT along the optical axis using a pump power of $P_{\text{pu}}=4\,$mW and it is the origin of the offset visible in the time-resolved Faraday rotation shown in Suppl.~Fig.~\ref{fig1}(b). The initial signal variations after each pump pulse are related to the optically excited hole in trion which recombines on the timescale of $0.4\,$ns~\cite{greilichPRB06,Greilich2006a}. Note also that no oscillatory component is present such that we can safely assume that the external field is oriented strictly along the $z$~axis. Any external field component perpendicular to the optical axis would lead to a precession of the electron spin about this component.

\begin{figure}[t]
	\begin{center}
		\includegraphics[trim=0mm 0mm 0mm 0mm, clip, width=\columnwidth]{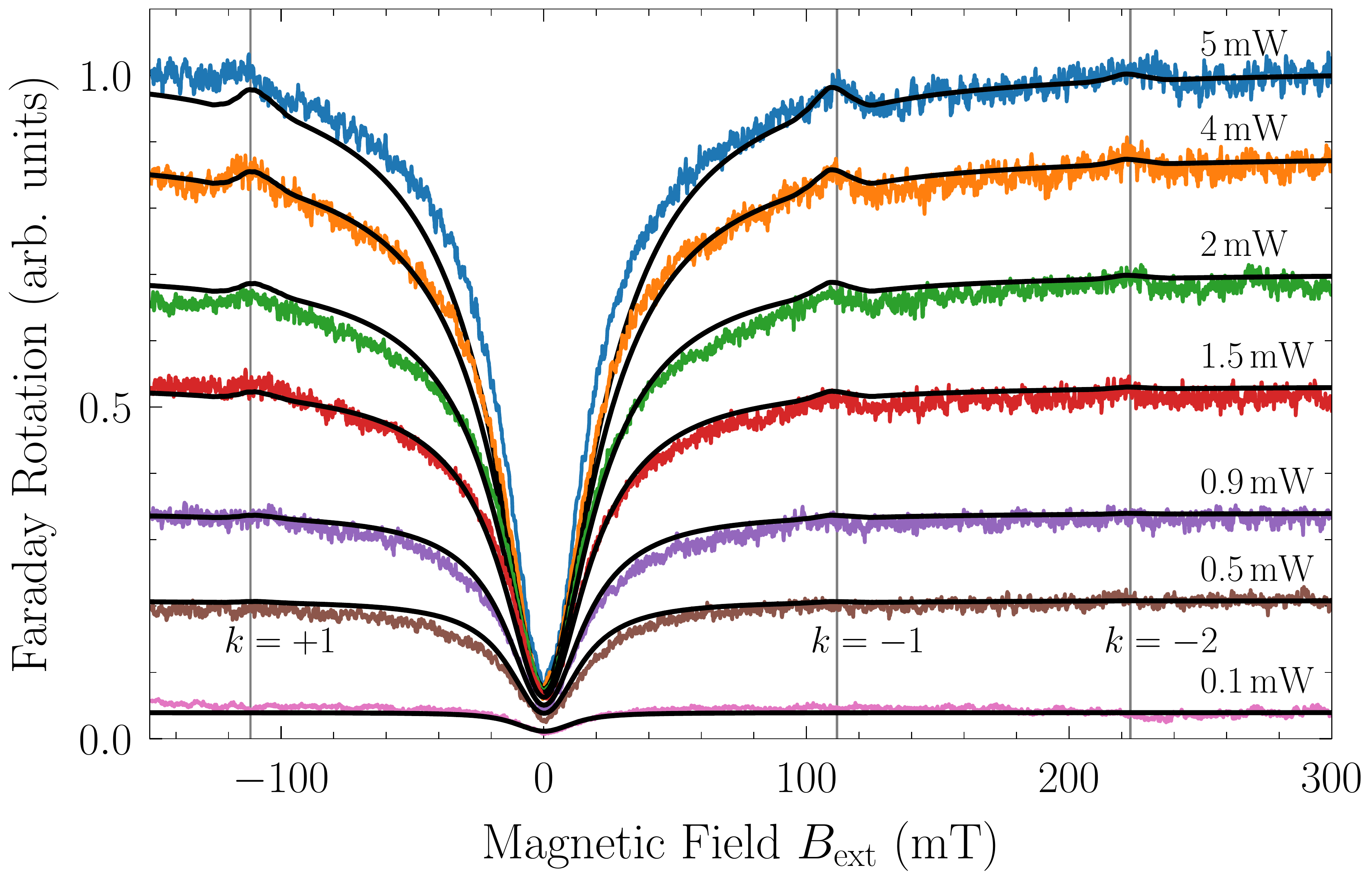}
		\caption{PRCs for different pump powers given next to the curves. The colored data is the experimentally measured Faraday rotation, the black curves are the simulated ones. The simulated PRCs are smoothed by a moving average over a range of $7\,$mT, the pulse area in the simulations is chosen to fit the experimentally obtained power dependence of the saturation amplitude~$A$, see Suppl.~Fig.~\ref{fig3}(a). The positions of the small peaks around $\Bext=-112\,$mT, $112\,$mT, and $224\,$mT match the RSA modes (vertical grey lines) predicted by the PSC~\eqref{PSC} for $k=+1$, $-1$, and $-2$, yielding $g_z = -0.64 \pm 0.01$ for the longitudinal electronic $g$~factor.}
		\label{fig2}
	\end{center}
\end{figure}

The polarization recovery curves~(PRCs) studied in this section are measured by setting the delay between subsequent pump and probe pulses to $-50\,$ps, i.e., the Faraday rotation is measured right before the arrival of the next pump pulse after a delay of $950\,$ps as indicated by the orange arrow in Suppl.~Fig.~\ref{fig1}(b). The longitudinal field is varied from $-150$ to $300\,$mT with a rate of~$45\,$mT$\,\text{min}^{-1}$. 

Supplementary~Figure~\ref{fig2} shows the PRCs for a wide range of pump powers. They have a $\mathsf{V}$-like typical for $n$-doped QDs~\cite{Petrov2008,Zhukov2018,Smirnov2018}.
For small pump powers, the electron spin polarization is minimal at $0\,$mT and rises to a saturation level within $30\,$mT. Without a magnetic field, the electron spin is only subject to the isotropic nuclear fluctuation field described by the characteristic frequency~$\wn$, leading to nuclei-induced spin relaxation.
For rather small fields, the electronic Zeeman effect does not dominate over the hyperfine interaction between the electron spin and nuclear spins. For larger fields, the external field dominates the nuclear fluctuation field and eventually, no nuclei-induced spin relaxation takes place. This results in a plateau region for the spin polarization~\cite{Glazov_book}.

\begin{figure}[t]
	\begin{center}
		\includegraphics[trim=0mm 0mm 0mm 0mm, clip, width=\columnwidth]{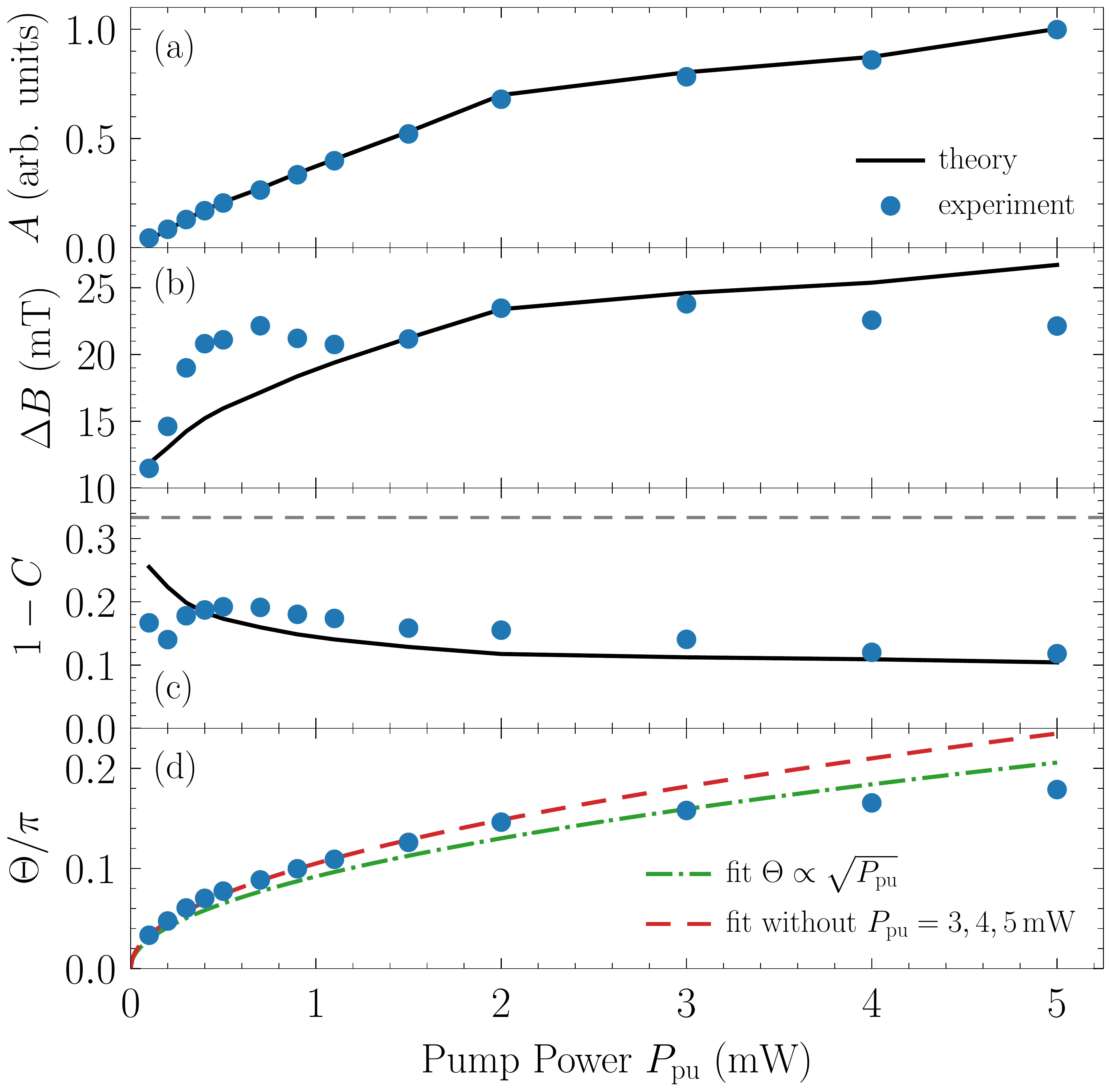}
		\caption{Characterization of the PRCs for experiment and theory. The saturation level~$A$, the width~$\DeltaB$ of the zero-field minimum, the ratio~$1-C$ which describes the depth of the zero-field minimum relative to $A$, and the pulse area~$\Theta$ are shown versus the pump power $P_\mathrm{pu}$. The pulse area is set in the model such that it fits the saturation level extracted from the experiment; the parameters $A$, $\DeltaB$, and $C$ are determined by fitting~\eqref{eq:PRC_fit} to the PRCs. The horizontal dashed line in~(c) marks the expected ratio $1-C = 1/3$ for small powers. For small powers, the pulse area shown in~(d) scales like $\Theta \propto \sqrt{P_\mathrm{pu}}$.}
		\label{fig3}
	\end{center}
\end{figure}

We characterize the measured and simulated PRCs by applying the fit function~\cite{Petrov2008}
\begin{align}
	R(\Bext) = A \left[ 1 - \frac{C}{1 + (\Bext-B_0)^2/\DeltaB^2}\right] \,.	\label{eq:PRC_fit}
\end{align}
The parameter $A$ defines the saturation level for large magnetic fields, $1-C$ is the ratio of spin polarizations for zero and large field, $\DeltaB$ characterizes the width of the PRC, and $B_0$ accounts for a potential shift.
The width of the PRC for small pump power is determined by the nuclear fluctuation field and can be extracted from the data by extrapolating the width to zero pump power, see~Suppl.~Fig.~\ref{fig3}(b). 
It amounts to an effective Overhauser field characterized by $\wn/(2\pi) = 70\,$MHz (equivalent to a field of $7.8\,$mT) for this sample~\cite{Zhukov2018}.
Supplementary~Figure~\ref{fig3}(a) shows the increase of the saturation level~$A$ as a function of the pump power~$P_\mathrm{pu}$.
For strong pulses corresponding to a larger pump power, a larger magnetic field is required to suppress the nuclei-induced spin relaxation~\cite{Schering2019,Smirnov2020b}, leading to an increased width of the PRC as shown in Suppl.~Fig.~\ref{fig3}(b).
The ratio~$1-C$ deviates from the standard value~$1/3$~\cite{Petrov2008}, which is only valid in the limit of very weak pulses~\cite{Schering2019,Smirnov2020b}, see Suppl.~Fig.~\ref{fig3}(c).

Very similar to the PRCs presented in the main text, additional modulations of the electron spin polarization appear in the PRCs at certain values of the longitudinal magnetic field in the case of larger pump powers.
This is the result of RSA in Faraday geometry due to nuclear fluctuation fields~\cite{Schering2019}. 
As shown in Suppl.~Fig.~\ref{fig2}, the modulations appear at magnetic fields $\Bext$ which fulfill the phase synchronization condition~(PSC)~\cite{Schering2019}
\begin{align}
	\Omega_\mathrm{L}= |k| \wR \,, \qquad k \in \mathds{Z} \,, \label{PSC}
\end{align}
for the Larmor frequency $\Omega_\mathrm{L} = \muB |g_z \Bext| \hbar^{-1}$ ($\muB$~is the Bohr magneton and $\hbar$~the reduced Planck constant).
These discrete resonance frequencies are given by multiples of the laser repetition frequency~$\wR$. We label them by the mode number~$k$.
The longitudinal electronic $g$~factor $|g_z|$ is the relevant system parameter which fixes the mode positions. By determining the maxima of these modulations, we obtain $|g_z| = 0.64 \pm 0.01$. From earlier experiments we know that in (In,Ga)As/GaAs QDs the $g$~factor has a negative sign~\cite{Yugova2007}, i.e., $g_z = -0.64 \pm 0.01$.

The impressive agreement between theory and experiment is achieved despite having only three fit parameters: the effective pulse area $\Theta$ (order of magnitude estimated), the spin relaxation time of the trion $\taust = 0.06\,\mu$s, and the longitudinal electronic $g$~factor $g_z = -0.64$. The remaining parameters are taken from the earlier sample characterization presented in Ref.~\cite{Zhukov2018}; they are listed in Table~\ref{tab:parameters}.

\begin{figure}[t]
	\begin{center}
		\includegraphics[trim=0mm 0mm 0mm 0mm, clip, width=\columnwidth]{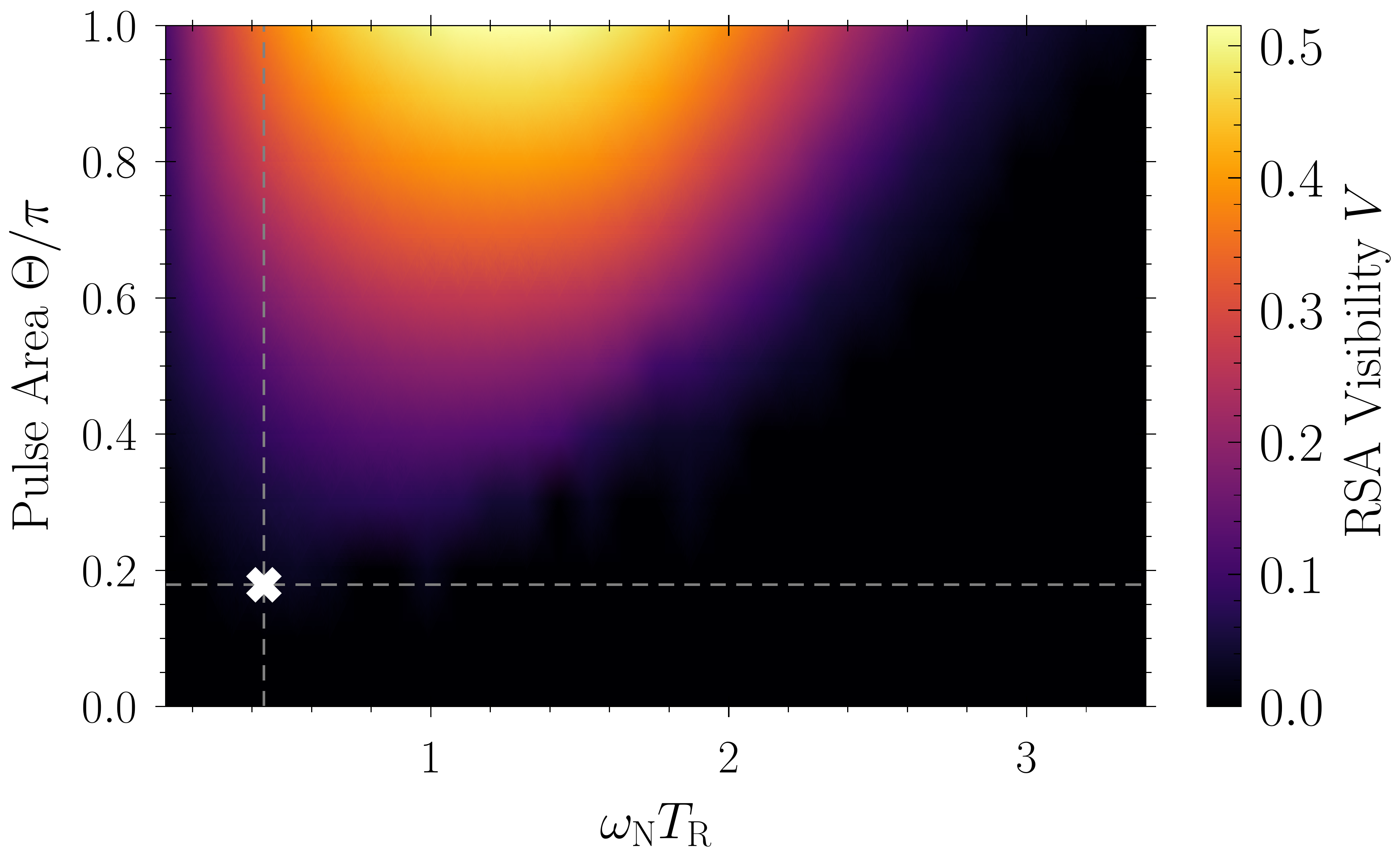}
		\caption{Visibility map of the RSA modes in Faraday geometry modeled for detection by Faraday rotation. The pump-probe delay is set to zero. The white cross marks the experimental conditions for a pump power of $P_\mathrm{pu} = 5\,$mW ($\Theta \approx 0.18\pi$) and a pulse repetition period of $\TR = 1\,$ns [\mbox{$\wn/(2\pi) = 70\,$MHz}].}
		\label{fig4}
	\end{center}
\end{figure}

Analogously to the main text, we benchmark our laser source by studying the RSA visibility~\cite{Smirnov2020b}
\begin{align}
	V \coloneqq \frac{R_\mathrm{max} - R_\mathrm{min}}{R_\mathrm{max}} \,,
\end{align}
where $R_\mathrm{max}$ is the Faraday rotation of the first maximum at the RSA mode $|k|=1$ and $R_\mathrm{min}$ denotes the adjacent minimum at a larger field $|\Bext|$. 
Note that the Faraday rotation is more difficult to calculate numerically than the Faraday ellipticity. The reason is a sign problem arising in the applied Monte Carlo sampling which results in much larger statistical fluctuations than for the Faraday ellipticity studied in the main text.
Since these fluctuations hinder the reliable detection of physical minima and maxima in the simulated PRCs, we set \mbox{$V = 0$} whenever $V < 0.02$.
A map of the RSA visibility in dependence of the pulse area~$\Theta$ and of the product of repetition period~$\TR = 2\pi/\wR$ and nuclear fluctuation field~$\wn/(2\pi) = 70\,$MHz is shown in Suppl.~Fig.~\ref{fig4}. The visibility becomes larger upon increasing the pulse area, which corresponds to a larger pump power. For a too large or too small repetition period, the visibility vanishes. Observing RSA in Faraday geometry when measuring the Faraday rotation appears to be easiest in the intermediate regime $\wn \TR \sim 1$.
As evident from the white cross on the color map, the laser source used in the experiment does not operate at optimal conditions. For a pump power of $P_\mathrm{pu}$, the visibility amounts to only $V \approx 0.02$. Under better conditions, the visibility could reach up to $V \approx 0.5$. 
No RSA modes are visible for the commonly used repetition periods $\TR = 13.2$~and~$6.6\,$ns ($\wn\TR \approx 5.8$ and $2.9$) in theory nor in experiment (not shown here).

The visibility for this QD sample with \mbox{$\wn/(2\pi) = 70\,$MHz} could be improved by increasing the pump power and by using a slightly larger repetition period of about~$2-3\,$ns.
A larger repetition period would also increase the number of visible RSA modes, allowing for an even more accurate determination of the longitudinal electronic $g$~factor.
In contrast, for sample~A analyzed in the main text~[\mbox{$\wn/(2\pi) = 140\,$MHz}] the choice $\TR=1\,$ns is perfectly suited to reveal the RSA modes because the nuclear spin fluctuations characterized by~$\wn$ are larger by a factor of 2.
The larger value also results in broader PRCs such that RSA modes are also visible at larger magnetic fields.
Our theoretical modeling reveals that measuring the Faraday ellipticity instead of the Faraday rotation for a degenerate pump-probe setup yields a larger RSA visibility. Hence, we opted for it in the experiments presented in the main text.
Another possibility to enhance the visibility is to exploit the spin-inertia effect; see the main text for details.

Generally, the RSA visibility strongly depends on the average spin polarization, which in turn depends on the spin relaxation time~$\taust$ of the hole in the intermediate trion, with a smaller time corresponding to a larger polarization~\cite{Smirnov2018}. As discussed in the main text, approaching the saturation limit of the spin polarization is detrimental for RSA in Faraday geometry, and hence, the RSA visibility is fairly sensitive to the choice of~$\taust$.


%